\documentclass{iau} 
\usepackage{graphicx,hyperref,natbib,amsmath}

\title{Schwarzschild modeling of barred galaxies}

\author[Vasiliev \& Valluri]{Eugene Vasiliev$^{1,2,3}$, Monica Valluri$^3$}

\affiliation{$^1$Institute of Astronomy, University of Cambridge, UK\\[\affilskip]
$^2$Lebedev Physical Institute, Moscow, Russia\\[\affilskip]
$^3$Department of Astronomy, University of Michigan, Ann Arbor, MI, USA\\[\affilskip]
email: {\tt eugvas@lpi.ru}
}

\pubyear{2019}
\volume{353}
\jname{Galactic Dynamics in the Era of Large Surveys}
\editors{M. Valluri \& J. Sellwood}

\renewcommand{\d}{\mathrm{d}}

\newcommand{\Bx}{\boldsymbol{x}}
\newcommand{\Bv}{\boldsymbol{v}}

\begin{document}

\maketitle

\begin{abstract}
We review the Schwarzschild orbit-superposition approach and present a new implementation of this method, which can deal with a large class of systems, including rotating barred disk galaxies.
We discuss two conceptuals problems in this field:
the intrinsic degeneracy of determining the potential from line-of-sight kinematics, and the   non-uniqueness of deprojection and related biases in potential inference, especially acute for triaxial bars. When applied to mock datasets with known 3d shape, our method correctly recovers the pattern speed and other potential parameters. However, more work is needed to systematically address these two problems for real observational datasets.
\end{abstract}

\firstsection\vspace{-2mm}
\section{Introduction}

A large fraction (1/3 to 2/3) of disk galaxies have bars. They are apparent in surface density profiles of face-on and moderately inclined galaxies, and have specific kinematic signatures which can be detected even in unfavourable (close to edge-on) orientations, e.g., a positive correlation between mean velocity and the third Gauss--Hermite moment $h_3$ \citep[e.g.][]{Bureau2005,Li2018}. Since stellar kinematics is widely used to constrain the gravitational potential, and in particular, measure the masses of central supermassive black holes (SMBH), the presence of a bar may affect the results of stellar-dynamical modelling, if not taken into account properly \citep{Brown2013}.

On the other hand, barred galaxies have complicated morphology, often with features such as boxy/peanut shape, long and short bar components, etc., and similarly complex orbital and kinematic structure. They require equally sophisticated modelling approaches, and at present there are two classes of methods which have been used for this task. Models of the Milky Way bulge/bar have been constructed with the Schwarzschild orbit-superposition method \citep{Zhao1996} and the made-to-measure (tailored $N$-body) method \citep{Portail2015, Hunt2013, Long2013}. Dynamical studies of external galaxies with bars are still in early stages.

In this contribution, we present a new implementation of the Schwarzschild method suitable to construct models of barred galaxies constrained by kinematic observations, and review several important conceptual and practical difficulties in this endeavor.

\vspace{-2mm}
\section{Schwarzschild method in a nutshell}

The problem of designing a gravitationally self-consistent model can be formulated as follows.
The Jeans theorem states that if the gravitational potential $\Phi(\Bx)$ and the distribution function (DF) of stars $f(\Bx, \Bv)$ are both stationary, then $f$ may depend on the phase-space coordinates only through the integrals of motion $\mathcal I(\Bx, \Bv\:|\:\Phi)$, although in practice the task of determining these integrals is not at all trivial in non-spherical potentials. The density $\rho(\Bx)$ of stars, on the one hand, is the 0th moment of DF:
\begin{equation}  \label{eq:rho_from_df}
\rho(\Bx) = \iiint f\big( \mathcal I[\Bx, \Bv\:|\:\Phi]\big)\,\d^3v ,
\end{equation}
and on the other hand, is related to the potential by the Poisson equation:
\begin{equation}  \label{eq:poisson}
\nabla^2\Phi(\Bx) = 4\pi \,G\,\rho(\Bx) ,
\end{equation}
with obvious generalization for multicomponent systems (e.g., stars + dark matter).

This system of coupled integro-differential equations can be attacked from two sides:
\begin{itemize}
\item From $\Phi$ to $f$: assume a particular functional form for $\Phi$ (and hence $\rho$) and find a suitable DF. Examples of this approach include the Eddington inversion formula or its anisotropic generalizations (e.g., \citealt{Cuddeford1991}).
\item From $f$ to $\Phi$: assume some functional form for $f$ and determine $\Phi$. In a spherical case, this approach leads to Plummer and King models, or their anisotropic generalizations. In non-spherical systems, one typically needs to follow an iterative procedure \citep[e.g.,][]{Kuijken1995,Binney2014}: starting from some initial potential, determine the integrals motion, then compute the density generated by the DF (\ref{eq:rho_from_df}), update the potential (\ref{eq:poisson}), and repeat until convergence.
\end{itemize}

A very general method for constructing equilibrium models was introduced by Martin \citet{Schwarzschild1979}. It belongs to the first family, namely constructs $f$ for the given density profile of arbitrary geometry. The problem of finding the integrals of motion is circumvented by by performing numerical orbit integration. Each orbit is essentially a $\delta$-function in the space of integrals, and the overall DF may be represented by a weighted sum of orbits, with weights hitherto undetermined:
\begin{equation}  \label{eq:df_schw}
f(\mathcal I) = \sum\nolimits_{k=1}^{N_\mathrm{orb}} w_k\, \delta(\mathcal I - \mathcal I_k).
\end{equation}
The density, in turn, is also discretized: defining a suitable spatial grid covering [almost] the entire density profile, one may compute the mass in each cell $\mathcal V_c$ of this grid:
\begin{equation}  \label{eq:cell_mass}
M_c \equiv \iiint_{x\in \mathcal V_c} \rho(\Bx)\, \d^3x.
\end{equation}
On the other hand, the contribution of each orbit to each cell (fraction of time $t_{kc}$ that $k$-th orbit spends in $c$-th cell) is also recorded during orbit integration. Then (\ref{eq:rho_from_df}) is reduced to a matrix equation for the vector of orbit weights:
\begin{equation}  \label{eq:matrix_eqn}
\sum\nolimits_{k=1}^{N_\mathrm{orb}} w_k\, t_{kc} = M_c, \quad c=1..N_\mathrm{cell}, \quad w_k \ge 0.
\end{equation}
Because of the non-negativity constraints on the orbit weights, this is not just a simple linear algebra problem, but an instance of so-called linear programming problems, in which an objective function is minimized w.r.t. parameters (in this case, orbit weights), subject to a set of linear inequality constraints. In the original paper \citep{Schwarzschild1979}, the particular choice of objective function was unimportant, the goal being simply to demonstrate the existence of a solution.

Soon it was realized that this method can be used to construct galactic models constrained by observations, namely the surface brightness profile and the line-of-sight velocity distribution (LOSVD) of stars. The latter is derived from the spatially resolved spectral datacubes and is usually parametrized as a Gauss--Hermite series \citep{Gerhard1993, vdMarel1993}, or in some other non-parametric form \citep[e.g., cubic spline,][]{Merritt1997}. Regardless of the parametrization, the contribution of $k$-th orbit to the observed LOSVD coefficient $U_n$ is given by another matrix $u_{kn}$, also constructed during orbit integration. Then the objective function to be minimized is the uncertainty-weighted deviation of the model from observational kinematic constraints plus a regularization term $\lambda\mathcal S$, whose role is to discourage unrealistically wiggly DFs:
\begin{equation}  \label{eq:objective}
\mathcal Q = \chi^2 + \lambda\mathcal S \equiv\sum_{n=1}^{N_\mathrm{obs}}  \Bigg( 
\frac{ \sum_{k=1}^{N_\mathrm{orb}} w_k\,u_{kn} - U_n }{\delta U_n} \Bigg)^2 + 
\lambda\mathcal S\big(\{w_k\}\big).
\end{equation}

One of the most common goals of dynamical modelling is to constrain the total potential of the galaxy, which is composed of the stellar (visible) component and the invisible components: dark matter halo, SMBH, etc. The stellar distribution is given by deprojecting the observed surface brightness profile (Section~\ref{sec:deprojection}) and multiplying it by the mass-to-light ratio, which is usually a free parameter in the model. For each choice of parameters for the total potential, a new orbit library needs to be constructed (unless the potential is simply scaled in amplitude, which is equivalent to the rescaling of velocity). Then a separate solution of the optimization problem (\ref{eq:matrix_eqn}, \ref{eq:objective}) is obtained, and the resulting values of $\chi^2$ for each model are used to determine the best-fit values and uncertainty intervals for the potential parameters.

Over the last two decades, several independent implementations of the Schwarzschild method have been developed. The most commonly used are the three axisymmetric codes: "Leiden" \citep{Rix1997,Cretton1999,Krajnovic2005}, "Nukers" \citep{Gebhardt2000,Thomas2004,Siopis2009}, "MasMod" \citep{Valluri2004}, and the triaxial code of \citet{vdBosch2008,vdVen2008,Zhu2018}. However, none of them are suitable for strongly flattened and rotating potentials of barred disk galaxies.

\section{The new Schwarzschild modelling code}  \label{sec:new_code}

We present a new implementation of the Schwarzschild method, which descends from the \textsc{Smile} code \citep{Vasiliev2013,Vasiliev2015}.
Its most important features are:
\begin{itemize}
\item It works in any geometry from spherical to axisymmetric to triaxial, and can deal with arbitrary density/potential profiles thanks to the flexible Poisson solvers.
\item It is possible to construct models of triaxial barred galaxies, which are stationary in the rotating reference frame.
\item Initial conditions for the orbit library are generated by randomly sampling the position from the density profile, and then assigning the velocity by one of the available ``seed'' methods (either Jeans equations or the Cuddeford--Eddington inversion formula).
\item It offers several choices for the density profile discretization, including piecewise-linear basis functions (as opposed to $\sqcap$-shaped cells in other codes).
\item The 3d ($X,Y,V_\mathrm{los}$) kinematic datacube is represented by 2nd or 3rd-degree tensor-product B-splines, which provide much higher accuracy than the commonly used histogramming approach (equivalent to 0th-degree B-splines).
\item Observational kinematic constraints can be provided either in the form of Gauss--Hermite moments, or a full LOSVD.
\item The code uses a very efficient quadratic optimization solver, which can routinely handle problems with $\mathcal O(10^4)$ constraints and $\mathcal O(10^5)$ orbits in only a few minutes on a modern multi-core workstation.
\item The code is publicly available as part of \textsc{Agama} library for dynamical modelling \citep{Vasiliev2019}: \url{http://agama.software}.
\end{itemize}

\section{Degeneracies in measurement of gravitational potential}  \label{sec:potential_constraints}

Consider a simple spherical model for stellar distribution around a central SMBH, for which the radial dependence of the density profile is known up to a normalization factor, and the only free parameters are the mass-to-light ratio of stars $\Upsilon$ and the mass of the SMBH $M_\bullet$.
If we only have access to the radial profile of the line-of-sight velocity dispersion $\sigma_\mathrm{los}(R)$, and use, e.g., Jeans equations to constrain the potential, it is well-known that this problem suffers from the so-called mass--anisotropy degeneracy \citep[e.g.,][Section 4.9.1]{BinneyTremaine}: the same $\sigma_\mathrm{los}(R)$ profile may be produced by different combinations of the potential and the radial profile of velocity anisotropy coefficient $\beta(r)$. To a large extent, this is caused by using only the 1d velocity dispersion profile instead of the full 2d LOSVD $\mathcal F (R, V_\mathrm{los})$.

\citet{Dejonghe1992} have shown that in a given potential $\Phi(r)$, the DF of a spherical system $f(E,L)$ can be uniquely computed from the LOSVD (or, more precisely, an infinite set of its moments). However, different choices of potential correspond to different inferred DFs: we would like to simultaneously compute the 2d function $f(E,L)$, together with the 1d function $\Phi(r)$, having only one 2d function $\mathcal F(R, V_\mathrm{los})$ at our disposal. In a more general (non-spherical) case, we would need to compute the DF as a function of three integrals of motion, together with the 3d potential, from just one 3d function $\mathcal F(X,Y,V_\mathrm{los})$, so the problem is still underdetermined. The only strong constraints on the potential come from the requirement that the computed DF should be non-negative. \citet{Dejonghe1992} conjectured that this requirement can be satisfied only by a narrow range of potentials, but have not demonstrated it rigorously. 

\begin{figure}
\includegraphics{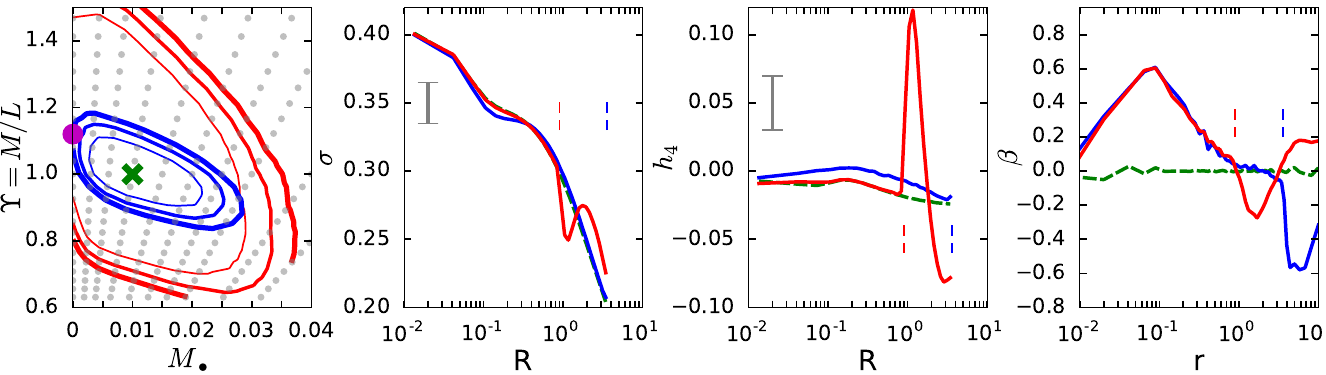}
\caption{
Results of fitting a two-parameter model ($\Upsilon={}$mass-to-light ratio, $M_\bullet={}$mass of the central SMBH) to a spherical model (a Hernquist profile with unit scale radius and unit mass, with a black hole of $M_\bullet=0.01$).
Left panel: contours of $\chi^2$ in the parameter space; red (outer) contours show the $1,2,3$-$\sigma$ regions for a model constrained by kinematic data out to $R=0.5\,R_\mathrm{half-light}=0.9$ (marked by a vertical red dashed line in the remaining panels), blue (inner) contours -- for a model with constraints extended up to $R=2\,R_\mathrm{half-light}$ (marked by a blue dashed line). The true values of $\Upsilon$ and $M_\bullet$ are marked by green cross, and kinematic profiles for this model are plotted by a green dashed line in the remaining panels (parameters of the Gauss--Hermite expansion -- $\sigma$, $h_4$, and the velocity anisotropy coefficient $\beta$). Examples of models without a black hole are marked by a purple dot in the left panel, and shown by red and blue lines in the remaining panels. The ``red'' model with kinematic constraints extending to a smaller radius is, of course, more flexible, and can fit the kinematics perfectly even for a wrong (zero) $M_\bullet$, but at the expense of a sudden change in intrinsic and projected velocity distributions just outside the radius where the kinematic constraints are provided. The ``blue'' model is a worse fit (most noticeably in the $h_4$ profile), but in a larger range of radii.
}  \label{fig:sphericalfit}
\end{figure}

Figure~\ref{fig:sphericalfit} shows the result of such an experiment performed with the Schwarzschild modelling approach. If the kinematic constraints have a limited spatial extent (red curves: up to $0.5\,R_\mathrm{half-light}$), a large range of $\Upsilon$ and $M_\bullet$ are equally consistent with the measured LOSVD profiles, parametrized by the first 6 Gauss--Hermite moments (as in the state-of-the-art observational datasets). If we increase the spatial coverage to $2\,R_\mathrm{half-light}$, then the allowed range of $\Upsilon$ is considerably narrower, and the range of allowed $M_\bullet$ also shrinks somewhat; however, the models with half or twice the true SMBH mass are still fit perfectly. In this case, we had a very optimistic choice of parameters (relatively large $M_\bullet$, and the PSF width of only 0.03, three times smaller than the influence radius); in realistic situations, the radius of influence is often barely, if at all, resolved. Of course, best-fit models with different parameters have substantially different internal kinematic structure (e.g., the velocity anisotropy parameter $\beta$, shown on the right panel of Figure~\ref{fig:sphericalfit}); moreover, it may vary quite dramatically outside the range of radii constrained by the data, calling for some additional priors or regularization constraints, which are outside the scope of this study.

The above experiment reiterates the conclusion reached in \citet{Valluri2004} that a large range of models can perfectly fit the same noiseless kinematic data. Such an idealized setup is, of course, not encountered in practice, and one must consider the results of fitting to noisy data. In this case, even the inherently flexible Schwarzschild method cannot fit the data perfectly (and, of course, should not attempt to fit all the wiggles caused by noise), and $\chi^2$ as a function of parameters of the potential near their best-fit values is closer to a parabolic function rather than a flat-bottomed trough, as discussed by \citet{Magorrian2006}. 

The uncertainty intervals on model parameters also shrink in the realistically noisy case, often up to the point of becoming implausibly small. The open problem is, though, that if one believes that there \textit{is} some intrinsic degeneracy in noise-free models, then a statistically sound method should be able to recover this degeneracy even in the presence of noise. One promising way to achieve this is bootstrapping in the space of orbit libraries: using a randomly chosen subset of orbits in the fitting procedure, determine the best-fit parameters of the potential, and then repeat this many times for different subsets of orbits. The distribution of best-fit parameters may better describe the real uncertainty intervals than the $\chi^2$ contours for the same set of orbits. In some sense, this approach is complementary to the suggestion of \citet{Magorrian2006} to marginalize over all possible DFs for a given potential, instead of considering just the best-fit one. This marginalization is extremely costly or even nearly infeasible in the case of Schwarzschild models, although \citet{Bovy2018} recently presented a proof-of-concept of this approach in the context of made-to-measure models. Substantially more work is needed to understand the caveats of fitting intrinsically degenerate models to noisy data, and their implications for determining reliable uncertainty intervals on potential parameters.

\section{Uncertainties and biases in deprojection}  \label{sec:deprojection}

\begin{figure}
\begin{tabular}{ccc}
\parbox[c]{1.7in}{\includegraphics[width=1.7in]{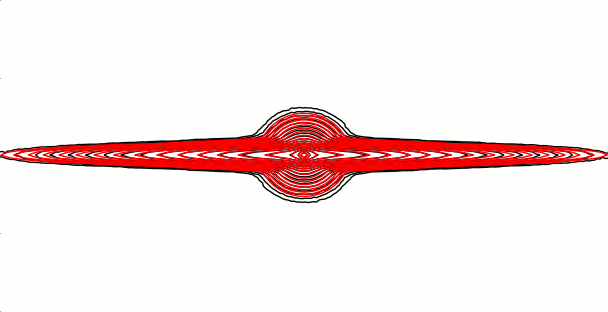}} &
\parbox[c]{1.7in}{\includegraphics[width=1.7in]{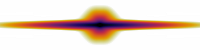}} &
\parbox[c]{1.7in}{\includegraphics[width=1.7in]{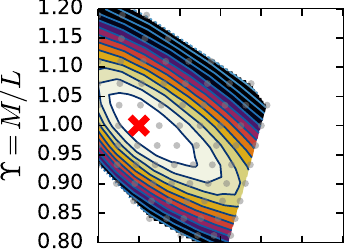}} \\
\parbox[c]{1.7in}{\includegraphics[width=1.7in]{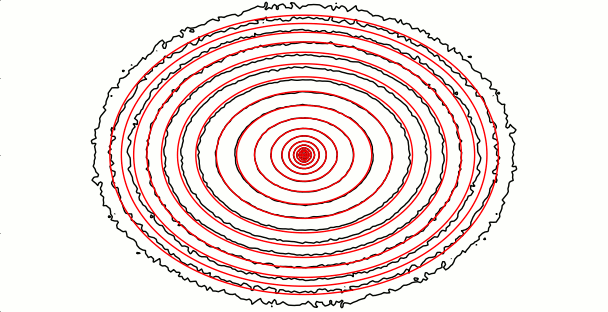}} &
\parbox[c]{1.7in}{\includegraphics[width=1.7in]{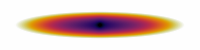}} &
\parbox[c]{1.7in}{\includegraphics[width=1.7in]{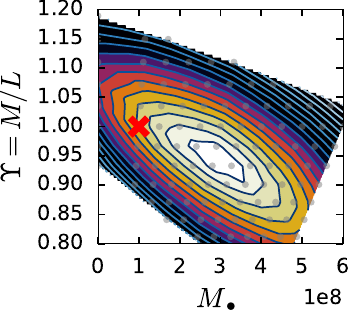}}
\end{tabular}
\caption{Example of biases in dynamical modelling introduced by deprojection.\protect\\
Top row shows an axisymmetric two-component (disk+bulge) galaxy seen edge-on, in which case the deprojection is unique. The results of Schwarzschild modelling with two free parameters (mass-to-light ratio $\Upsilon$ and the mass of the central SMBH $M_\bullet$) are shown in the right column: the true values are marked by red cross, and the contours of $\chi^2$ demonstrate that these can be recovered well (with the innermost contour enclosing models with essentially perfect fits to kinematics), although with a relative large uncertainty on $M_\bullet$.\protect\\
In the bottom row, the same galaxy observed at an inclination $45^\circ$ is fitted by an MGE (left panel), which is then deprojected assuming the true inclination angle, resulting in a noticeably thicker system without any pronounced bulge (middle panel). The best-fit parameters are now substantially biased compared to the true ones (right column).
}  \label{fig:deprojection}
\end{figure}

To construct orbit-superposition models, one needs to specify the 3d density of the system, but in practice we only observe the 2d surface brightness profile. The problem of inferring the 3d (intrinsic) density profile from the 2d (projected) one is known to be underdetermined, i.e., it has no unique solution (except spherical and edge-on axisymmetric cases, \citealt{Kochanek1996, Gerhard1996}).

A common approach to sidestep this problem is to assume that the 3d density profile is axisymmetric and ellipsoidally-stratified, i.e., the equidensity surfaces are ellipsoidal with constant axis ratios. In this case, the projected profile is also elliptically-stratified, and there is one-to-one correspondence between the intrinsic and projected axis ratio for a given (assumed) inclination angle. This approach is readily generalized to the multi-component case, where each projected elliptically-stratified component can be uniquely deprojected at the given inclination. The Multi-Gaussian Expansion (MGE) formalism \citep{Emsellem1994, Cappellari2002} is widely used to represent a broad range of galactic morphologies with $\mathcal O(10)$ Gaussian components. It can also be applied to triaxial systems, in which the orientations of individual projected ellipses are not the same, but the intrinsic ellipsoidal components are forced to be aligned.

Unfortunately, when the assumption of ellipsoidal shape is not satisfied (e.g., in the case of boxy bars or ``disky disks''), this method (or, in fact, any other similar approach) can lead to incorrect inference on the 3d shape of the system, as illustrated in Figure~\ref{fig:deprojection}.
Moreover, this may bias the results of dynamical modelling in a way that is hard to control or compensate for. The problem is likely even more severe for triaxial systems (including barred galaxies).

A detailed investigation of deprojection uncertainties and caveats is left for a future study, and in testing the Schwarzschild code on mock datasets, we assume a perfect knowledge of the shape of the 3d density profile.

\section{Measurement of bar pattern speed}

\begin{figure}
\includegraphics{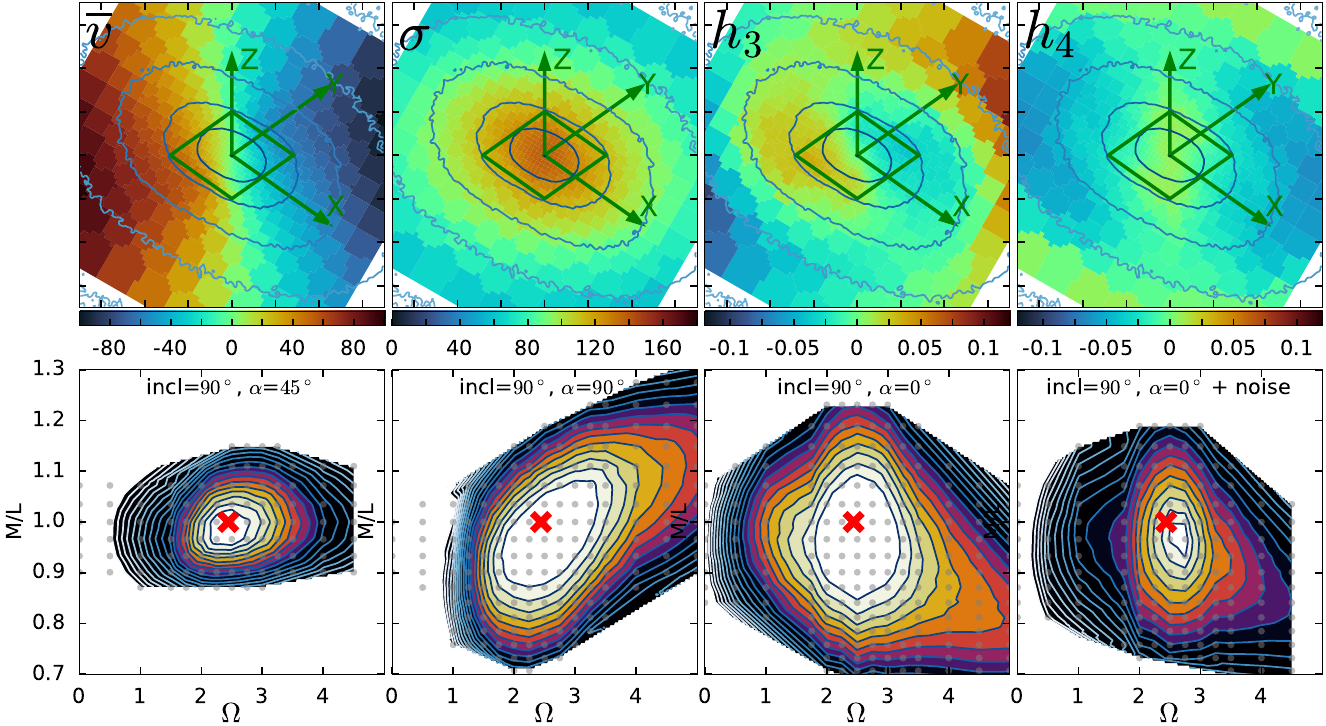}
\caption{
Top row: kinematic maps of a barred disk galaxy ($v, \sigma, h_3, h4$), with overplotted contours showing the surface density. The inclination is $45^\circ$ and the bar is rotated by $45^\circ$ from the line of nodes; the equatorial plane is shown by a green rectangle.\protect\\
Bottom row: likelihood surface as a function of two free parameters in Schwarzschild models: pattern speed $\Omega$ and mass-to-light ratio $\Upsilon$. Red cross indicates the true values; contours and color show the $\chi^2$ values for each model: white area has $\chi^2 \approx 0$ (except the rightmost panel), and subsequent contour level show $1\sigma$, $2\sigma$, $3\sigma$, \dots uncertainty regions for 2 degrees of freedom ($\delta\chi^2=2.3, 6.2, 11.8, \dots$). The leftmost panel corresponds to the model shown in the top row. The remaining panels show models observed at different orientations and hence having different kinematic maps (not shown here): edge-on (inclination $90^\circ$) and bar seen end-on ($\alpha=90^\circ$) or sideways ($\alpha=0$). The rightmost panel corresponds to the models fitted to kinematic maps with a realistic amount of added noise (5~km/s for $\overline v$ and $\sigma$, and 0.05 for Gauss--Hermite moments).
}  \label{fig:patternspeed}
\end{figure}

To demonstrate the correctness of the Schwarzschild code, we apply it to the problem of measuring the pattern speed of a barred disk galaxy. We use the $N$-body simulation from \citet{Fragkoudi2017}, with $10^7$ particles for the stellar component, embedded in a live dark matter halo (for which we assume a fixed parametric profile). Figure~\ref{fig:patternspeed} shows the noise-free kinematic maps of $v, \sigma, h_3, h_4$ (top row) of the system in the ``optimistic'' case ($45^\circ$ inclination with the bar rotated by $45^\circ$ from the line of nodes, producing a clear photometric misalignment and kinematic twist). The bottom left panel shows the contours of $\chi^2$ as a function of the mass-to-light ratio $\Upsilon$ and the pattern speed $\Omega$. The model with the true values of parameters is essentially a perfect fit, and even a 10\% deviation in $\Upsilon$ or a 20\% deviation in $\Omega$ is noticeably worse. Other panels in the bottom row show the results of the fit for models observed at less favorable orientations (assuming that the orientation is known); they demonstrate that the constraints are weaker but still unbiased, and even with a realistic amount of noise the true parameters are recovered remarkably well.

\section{Summary and outlook}

The problem of dynamical modelling of barred galaxies has several conceptual difficulties. Non-uniqueness of deprojection is especially troublesome for bars, which do not have ellipsoidally stratified density profiles. Incorrect assumptions about the intrinsic shape may lead to biases in dynamical inference on the potential. This adds to a more general problem of intrinsic degeneracy of recovering the potential and the DF simultaneously from a noisy 3-dimensional kinematic datacube.

On the other hand, the new implementation of the Schwarzschild method correctly recovers the potential and the pattern speed of mock galaxy models, when using the true shape of the density profile. The code is designed to be very flexible and efficient, and made available to the community. More work is needed to develop sufficiently general deprojection methods and reliably estimate the uncertainties on the potential parameters.

EV acknowledges support from the European Research council under the 7th framework programme (grant \#308024). MV acknowledges support from HST-AR-13890.001, US NSF award AST-1515001.


\end{document}